\documentclass{IEEEcsmag}
\usepackage[utf8]{inputenc}
\usepackage{amsmath, amssymb}
\DeclareMathOperator*{\argmin}{arg\,min}

\usepackage{cite}
\usepackage{graphicx}
\usepackage{url}
\usepackage{xcolor}
\usepackage{siunitx}
\usepackage{gensymb}
\usepackage{textcomp}
\usepackage[numbers,sort&compress]{natbib}
\usepackage{amsmath}
\usepackage{bm}
\usepackage{multirow,tabularx}
\usepackage{threeparttable}
\usepackage{makecell}
\usepackage{array}
\usepackage{graphicx}
\usepackage{subfig}
\newcolumntype{P}[1]{>{\centering\arraybackslash}p{#1}}
\newcolumntype{M}[1]{>{\centering\arraybackslash}m{#1}}
\usepackage[colorlinks,urlcolor=blue,linkcolor=blue,citecolor=blue]{hyperref}
\expandafter\def\expandafter\UrlBreaks\expandafter{\UrlBreaks\do\/\do\*\do\-\do\~\do\'\do\"\do\-}
\usepackage{upmath,color}

\jvol{XX}
\jnum{XX}
\paper{8}
\jmonth{Month}
\jname{Publication Name}
\jtitle{Publication Title}
\pubyear{2024}

\begin{document}

\sptitle{Article Type: Description  (see below for more detail)}

\title{Tile-Weighted Rate-Distortion Optimized Packet Scheduling for 360$\degree$ VR Video Streaming}

\author{Haopeng Wang}
\affil{University of Ottawa}

\author{Haiwei Dong}
\affil{Huawei Canada and University of Ottawa}

\author{Abdulmotaleb El Saddik}
\affil{University of Ottawa and MBZUAI}


\begin{abstract}\looseness-1
A key challenge of 360$\degree$ VR video streaming is ensuring high quality with limited network bandwidth. Currently, most studies focus on tile-based adaptive bitrate streaming to reduce bandwidth consumption, where resources in network nodes are not fully utilized. This article proposes a tile-weighted rate-distortion (TWRD) packet scheduling optimization system to reduce data volume and improve video quality. A multimodal spatial-temporal attention transformer is proposed to predict viewpoint with probability that is used to dynamically weight tiles and corresponding packets. The packet scheduling problem of determining which packets should be dropped is formulated as an optimization problem solved by a dynamic programming solution. Experiment results demonstrate the proposed method outperforms the existing methods under various conditions.
\end{abstract}

\maketitle

\chapteri{W}ith the popularity of VR \cite{9927197}, {360\degree} video a.k.a. VR video is also attracting more attentions. Streaming high-quality VR videos faces challenges due to bandwidth requirements and varying network conditions. However, a user only sees the contents inside the viewport at a time, and too much resource is wasted in delivering the rest content that the user does not view. Therefore, tile-based adaptive bitrate streaming approaches are proposed to reduce data volume. VR video streaming rarely uses the computational resources of network nodes. Network congestion and rebuffering occur when VR video is streamed over an inadequate bandwidth. Network nodes drop packets randomly, which can affect reconstructed video unexpectedly. If certain less important packets are dropped, the quality of the reconstructed video could degrade in a controllable and limited manner. Additionally, dropping packets outside the viewport could reduce the distortion of the viewport. 

\begin{figure*}[htbp]
\centering
\includegraphics[width=0.99\textwidth]{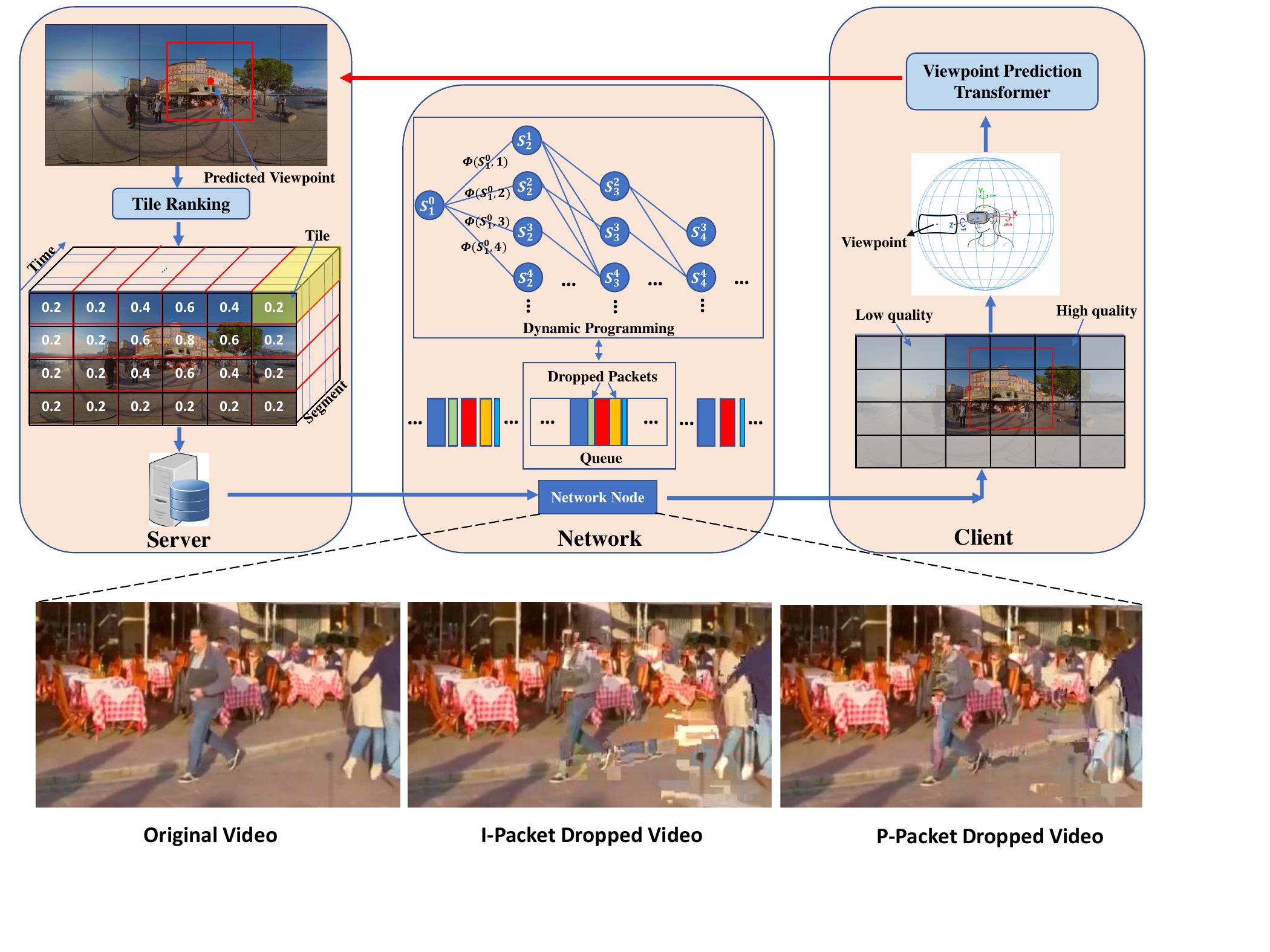}
\caption{Tile-weighted rate-distortion packet scheduling optimization system for VR video streaming. Each packet has a color indicating its importance, and a width indicating its size. Due to bandwidth restrictions, some packets are dropped according to the proposed method. The visual quality of the reconstructed video mainly depends on the type of the dropped packet.}
\label{view}
\end{figure*}

This article proposes a tile-weighted rate-distortion (TWRD) packet scheduling system to reduce data volume and video distortion based on a multimodal spatial-temporal attention transformer. A VR video is divided into tiles, where we assume each tile's frame has one packet \cite{https://doi.org/10.48550/arxiv.2202.04703, 7574700, 6172571}. The rate-distortion information of a packet consists of the rate and distortion, where the rate represents the size of the packet, and the distortion represents the quality impact if the packet is lost. As shown in Fig. \ref{view}, a VR video is initially streamed at the lowest quality. As the video plays, the transformer model predicts future viewpoints with probabilities based on the user's historical viewpoint trajectory. Every tile is then weighted based on the probability. When the queue size of a network node exceeds a threshold regardless of the bandwidth conditions and video quality, the network node starts the proposed dynamic programming solution to determine a packet scheduling scheme and drops packets as scheduled to avoid network congestion and reduce quality distortion. The distortion mainly depends on the type of dropped packets. As shown in Fig. \ref{view}, dropping an I-packet induces significant distortion that spans all frames within the GOP (group of pictures). Similarly, if a P-packet is lost, noticeable distortion occurs, making all subsequent frames incapable of decoding. Since other frames do not utilize B-packets, the drop of a B-packet results in the loss of only one reconstructed frame, which users cannot notice. Following is a summary of the contribution of this article:
\begin{itemize}
    \item A tile-weighted rate-distortion optimized VR video streaming system is proposed to improve video quality.
    \item A multimodal spatial-temporal attention transformer is proposed to predict viewpoint with probability that is used to rank tiles dynamically.
    \item The packet scheduling problem is formulated as an optimization problem which is solved by a proposed dynamic-programming-based solution. 
    \item Experiment results demonstrate that the proposed approach outperforms other approaches under varying bandwidths.
\end{itemize}

\begin{table*}[]
\caption{Comparison of the existing packet scheduling methods.}
\centering
\label{one}
\begin{threeparttable}
\begin{tabular}{|M{1.3cm}|M{1.2cm}|M{1.3cm}|M{0.95cm}|M{0.5cm}|p{4.5cm}|p{3.6cm}|M{1cm}|}
\hline
Citation    & Video Type  & Resolution  & Frame Number\tnote{1} & FPS\tnote{2} & \makecell[c]{Formulated Problem}    & \makecell[c]{Proposed Method}   &  Viewport-Aware in Network \\ \hline
Moharrami et al.    \cite{MOHARRAMI2023100149}              & Traditional video          & 1280 × 720 & ---  & 30  & Minimize quality degradation                       & Drop packets according to packet type  & No \\ \hline
Gobatto et al. \cite{https://doi.org/10.48550/arxiv.2202.04703}              & Traditional video          & 2560 × 1600 & 600  & 60  &  Minimize IRAP packet loss rate                       & Drop non-IRAP packets  & No \\ \hline

Chakareski et al. \cite{1608103} & Traditional video & 176 × 144   & 300 & 30  & Maximize overall quality of multiple videos streamed over a limited bandwidth transmission channel with rate-distortion information & Subgradient method with lagrangian relaxation to compute Lagrange multiplier of nonconstrained problem & No  \\ \hline
Corbillon et al. \cite{7574700}        & Traditional video & 1920 × 1080 & ---  & 25  & Minimize the degradation of video using an evaluation function considering frame type, dependencies, and size  & Drop frames according to their importance obtained from evaluation function   & No  \\ \hline
Li et al. \cite{4543842}        & Traditional video & 176 × 144 & 600  & 30  & Minimize the distortion of video using rate-distortion  & The problem is solved with a greedy algorithm   & No  \\ \hline
Nasralla et al. \cite{NASRALLA2018126} & Traditional video & 640 × 416   & 100  & 25  & Reduce packet delay and improve quality using a utility function based on temporal complexity and frame type   & Drop packets according to packets priority obtained from utility function        & No  \\ \hline
Change et al. \cite{6172571}           & Traditional video & 720 × 480   & 36000  & 30  & Minimize visual score indicating the visual impact of a frame loss   & Drop B-frame according to the visual score  & No \\ \hline
Cosma et al. \cite{9068423}            & VR video          & ---         & --- & --- & Improve the fraction of time in the transmission time interval by allocating the available frequency resources to different traffic classes  & Reinforcement learning with continuous actor-critic learning automata algorithm & No   \\ \hline

Chakareski \cite{9086630}              & VR video          & 3840 × 2048 & 450                                                            & 30  & Maximum VR video quality delivered from multiple base stations  considering content popularity, rate-distortion, and the information of base stations & The problem is solved with an approximate solution obtained using a faster iterative algorithm                  & No                                                                   \\ \hline
Ours                                                    & VR video          & 3840 × 1920 & 500                                                            & 25  & Minimize the distortion of the entire VR video and viewport over limited bandwidth transmission channel considering viewport and rate-distortion         & Dynamic programming containing state transition equation and initial state to solve the optimization problem     & Yes                                                                  \\ \hline
\end{tabular}%
\begin{tablenotes}
    \footnotesize               
        \item[1] Frame number represents the minimum total number of frames for a video.  
        \item[2] FPS represents frames per second.  
\end{tablenotes}
\end{threeparttable}
\end{table*}

\section{Related Work}
There are many existing packet scheduling algorithms for traditional video streaming. The tail drop algorithm is a simple and popular packet scheduling algorithm. However, the tail drop does not differentiate between packet types. Hence, dropping video frames based on frame types (I, P, B) with different priorities is proposed \cite{ MOHARRAMI2023100149}. Moharrami et al. \cite{MOHARRAMI2023100149} drop packets in the order of B, P, and I.
Gobatto et al. \cite{https://doi.org/10.48550/arxiv.2202.04703} preemptively drop non-IRAP packets. These methods, however, do not take into account differences between frames of the same type, as frames of the same type have different impacts on video quality. Therefore, more sophisticated methods are proposed \cite{1608103, 7574700, 8836788,4543842,NASRALLA2018126,6172571}. Chakareski et al. \cite{1608103} addressed the packet scheduling problem for multiple videos by characterizing video packets using rate-distortion information. However, they aim to achieve fairness between multiple videos. Corbillon et al. \cite{7574700} prioritize video packets using an evaluation function taking into account frame type, dependency, and size. Li et al. \cite{4543842} formulated the packet scheduling of traditional 2D video at the transmitter as an optimization problem to minimize distortion with rate-distortion information, and solved it with a greedy algorithm. Nasralla et al. \cite{NASRALLA2018126} proposed a content-aware packet scheduling method for video streaming. A utility function based on the temporal complexity and type of frames is proposed to prioritize packets. However, their work ignores the interdependencies of frames and the rate information. Chang et al. \cite{6172571} developed a visibility model for B-frame loss to generate a visual score for each frame before video transmission. However, only B-frame is allowed to be dropped in their system.

\begin{figure*}[htbp]
\centering
\includegraphics[width=1\textwidth]{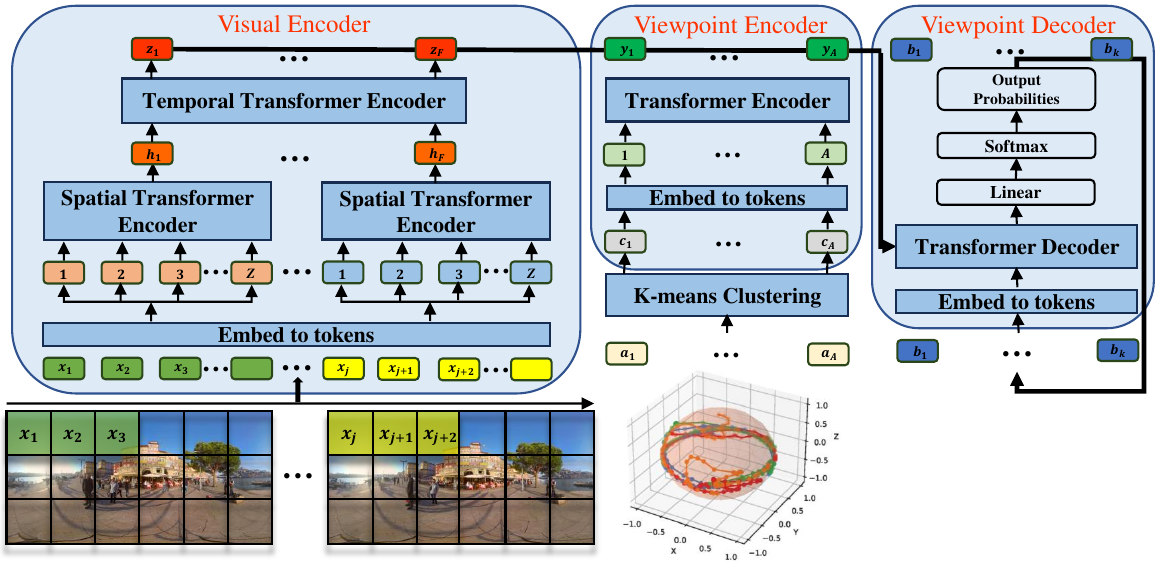}
\caption{The architecture of the multimodal spatial-temporal attention transformer model for viewpoint prediction with classification method.}
\label{transformer}
\end{figure*}

Despite the rapid development of VR video, very few packet scheduling methods are proposed. Cosma et al. \cite{9068423} proposed a packet scheduling method based on machine learning to allocate network resources for live VR video and other media applications instead of data reduction for a VR video. Chakareski \cite{9086630} integrates content popularity, rate-distortion and base station information to generate packet scheduling for resource allocation. The system, however, cannot realize a user's viewport in practice since it only uses a frequency model to determine the popularity of VR content.

Table \ref{one} compares the popular packet scheduling methods mentioned above for traditional and VR videos. Rather than reducing data volume in a network, existing VR video packet scheduling algorithms are mainly concerned with distributing resources among different traffic streams and multiple paths, which can not cope with variations of limited bandwidth. Meanwhile, existing packet scheduling methods for VR video cannot take account of viewport importance in the transmission network. Hence, we propose a content-aware packet scheduling strategy to reduce data volume and quality distortion.

\section{Viewpoint Prediction}

Given a set of historical viewpoints, $\{a_i\}^A_{i=1}$, where position $a_i \in \mathbb{R}^3$, and a sequence of video frames $\{f_j\}^F_{j=1}$, where frame $f_j \in \mathbb{R}^{H \times W \times C}$, and $H$, $W$ and $C$ are the height, width, and channel number. We aim to predict a viewer's viewpoint trajectory $\{b_r\}^{B}_{r=1}$. As shown in Fig. \ref{transformer}, the transformer contains a visual encoder, a viewpoint encoder, and a viewpoint decoder. The encoder-decoder architecture proposed in Vaswani et al.'s work \cite{10.5555/3295222.3295349} is used in our work. Each module is described next.

\subsection{Visual Encoder}
Each frame is split into $Z$ patches and patch $x_i \in \mathbb{R}^{h \times w}$, $i=1,...,Z$, where $h$ and $w$ are height and width. Each patch is embedded into tokens with patch and positional embedding, same as the ViT \cite{vit}. The spatial transformer encoder only models interactions between tokens extracted from the same frame. The spatial attention score of patch $x_i$ of frame $f_j$ is given by: 
\begin{equation}
    Attention_{spatial}=softmax(\frac{q_{(x_i,f_i)}}{\sqrt{d_k}} \cdot \{k_{(x_\alpha,f_j)}\}^T_{\alpha=1,...,Z})
\end{equation}
where $q_{(x_i,f_j)}$ and $k_{(x_\alpha,f_j)}$ are the query vector of patch $x_i$ and key vector of patch $x_\alpha$ in frame $f_j$, respectively. The $d_k$ is the embedding dimension.
A representation for each frame index $h_i$ is obtained after the spatial encoder. The frame-level representations, $h_i \in \mathbb{R}^{d_k} $, are concatenated into, $H \in \mathbb{R}^{F \times {d_k}}$, and then forwarded through a temporal encoder consisting of $N$ transformer layers to model interactions between tokens from different temporal frames. The temporal attention score of frame $f_j$ is given by:
\begin{equation}
    Attention_{temporal}=softmax(\frac{q_{(f_j)}}{\sqrt{d_k}} \cdot \{k_{(f_\beta)}\}^T_{\beta=1,...,F})
\end{equation}
where $q_{(f_j)}$ and $k_{(f_\beta)}$ are the query vector of frame $f_j$ and key vector of frame $f_\beta$. 
After being processed by the temporal encoder, the output tokens $\{z_j\}_{j=1}^{F}$, $z_j \in \mathbb{R}^{d_k}$ of this encoder are then obtained.  
\subsection{Viewpoint Encoder}
Here, the trajectory prediction is treated as a classification problem rather than as a regression problem that predicts coordinates directly. The original position $a_i$ is classified by the K-means clustering algorithm into different groups to get centroids. The centroid $c_i$ is embedded into tokens with word embedding and position embedding. Similarly, the viewpoint encoder models the relationship between tokens extracted from the historical viewpoint trajectory. The output tokens $\{y_i\}_{i=1}^A$, $y_i \in \mathbb{R}^{d_k}$ are obtained after viewpoint encoder.
\subsection{Viewpoint Decoder}
The viewpoint decoder generates the viewpoint set $\{b_r\}^{B}_{r=1}$ by using the encoder embeddings, which are obtained by concatenating the visual and viewpoint tokens $\{z_j\}_{j=1}^{F}$ and $\{y_i\}_{i=1}^A$. At each autoregressive step $k$, the viewpoint decoder relies on a causal transformer decoder which cross-attends with the encoder outputs and self-attends with tokens generated in previous steps to generate a representation. Then the model predicts the next token in the trajectory with the representation.

\section{Tile Ranking}
Existing methods \cite{9024132} classify tiles based on visual overlap without considering viewpoint probability, resulting in constant weight and resource consumption despite varying viewpoint probabilities. Thus, we dynamically compute the tile's weight using perspective probability and drop more packets inside the viewport when the probability is low and vice versa. 
We categorize tiles into four classes based on ``perceptual importance'': class-4 for tiles entirely within the viewport (highest rank), class-3 for those in over half of the viewport, class-2 for tiles with less than half in the viewport, and class-1 for tiles outside the viewport. At time step $t$, we calculate the weight of each tile by:
\begin{equation}
    \lambda_t = \frac{E \cdot L}{M}
\end{equation}
where $E$ is the probability of prediction obtained from the transformer model. $M$ is the number of classes, which is 4 in our work and $L$ is the class level, $L=1,2,3,4$. For example, as shown in Fig.\ref{view}, the predicted viewpoint has a probability of 0.8. The weight of class-4 tile is computed by $\lambda_t=\frac{0.8 \cdot 4}{4}=0.8$ while the weight of class-3 tile is $\lambda_t=\frac{0.8 \cdot 3}{4}=0.6$.

\section{Packet Scheduling Problem Formulation}
Consider that a VR video is divided into $t$ tiles $\bm{V}=\{V^0, V^1,...,V^{t-1}\}$. All tiles have the same number of frames $n$. Therefore, the frame sequence of tile $\tau$, $\tau=0,1,...,{t-1}$, can be expressed by $V^{\tau}=\{p^{\tau}_0, p^{\tau}_1,...,p^{\tau}_{n-1}\}$. The frame $\eta$, $\eta=0,1,...,{n-1}$, of the tile $\tau$ can be defined as $p_{\eta}^{\tau}$. The packet loss distortion is measured by using the structural similarity index measure (SSIM). We calculate distortion between two packets $p_i$ and $p_j$ with $d(p_i,p_j)=1-SSIM$.

For tile $\tau$, a transmission subset with a length of $l$, $S^{\tau}=\{p^{\tau}_{s_0}, p^{\tau}_{s_1},...,p^{\tau}_{s_{l-1}}\}$, is selected from $V^{\tau}$ to transmit over the network channel. The reconstructed video sequence of $V^\tau$ because of $S^{\tau}$ is denoted as $C_{V^{\tau}}(S^{\tau})=\{{p^{\tau}_0}^{\prime}, {p^{\tau}_1}^{\prime},...,{p^{\tau}_{n-1}}^{\prime}\}$. Dropped frames are compensated using the previous frame concealment. The $C_{V^{\tau}}(S^{\tau})$ is constructed by replacing the dropped frame with the nearest previous neighboring frame in $S^{\tau}$, which is,
\begin{equation}
    {p^{\tau}_i}^{\prime} = p^{\tau}_{j=max(s), s\in \{s_0, s_1, ..., s_{l-1}\}, j\leq i}
    \label{e0}
\end{equation}

We assume that distortions caused by multiple packet losses are additive \cite{4543842}. The distortion $D(p_{\eta}^{\tau})$ can be calculated by adding up the distortions between each frame and the corresponding reconstructed frame, which is
\begin{equation}
\begin{split}
    &D(p_{\eta}^{\tau}) = \sum_{i=0}^{n-1}d(p^{\tau}_i,{p^{\tau}_i}^{\prime})\\
    &=d(p^{\tau}_{\eta},p^{\tau}_{j=max(s), s\in \{s_0, s_1, ..., s_{l-1}\}, j\leq \eta})+\sum_{i=0, i\neq \eta}^{n-1}d(p^{\tau}_i,{p^{\tau}_i}^{\prime}) 
    \label{e1}
\end{split}
\end{equation}

It shows that distortion of $p_{\eta}^{\tau}$ contains two parts. The first part indicates the distortion on $p_{\eta}^{\tau}$ itself without considering error propagation. Without considering error propagation, the first part can be written as
\begin{equation}
\begin{split}
    \sum_{i=0}^{n-1}d(p^{\tau}_i,{p^{\tau}_i}^{\prime})&=\sum_{i=0}^{n-1}d(p^{\tau}_{i},p^{\tau}_{j=max(s), s\in \{s_0, s_1, ..., s_{l-1}\}, j\leq i}) \\
    &=d(p^{\tau}_{\eta},p^{\tau}_{\eta-1})
\end{split}
\end{equation}
The second part is the sum of distortion on all frames except $p_{\eta}^{\tau}$ itself due to error propagation. We denote second part as $\Omega(p_{\eta}^{\tau})$. Note that $d(p^{\tau}_i,{p^{\tau}_i}^{\prime})=0$ for $i<\eta$, as there is no previous distortion before the loss of packet $p_{\eta}^{\tau}$.
Therefore, Eq. \ref{e1} can be rewritten as
\begin{equation}
\begin{split}
    &D(p_{\eta}^{\tau}) = \sum_{i=0}^{n-1}d(p^{\tau}_{i},p^{\tau}_{j=max(s), s\in \{s_0, s_1, ..., s_{l-1}\}, j\leq i}) + \Omega(p_{\eta}^{\tau})
    \label{e2}
\end{split}
\end{equation}

Given the $V^{\tau}$ and $S^{\tau}$, a dropped set with length $z$, $K^{\tau} = \{p_{k_0}, p_{k_1}, ..., p_{k_{z-1}}\}$ is determined. Therefore, the total distortion of dropped set $K^{\tau}$ can be expressed by
\begin{small}
\begin{equation}
\begin{split}
        &D(K^{\tau})=\sum_{\eta\in K^{\tau}}D(p_{\eta}^{\tau}) \\
        &=\sum_{\eta\in K^{\tau}}\sum_{i=0}^{n-1}d(p^{\tau}_i,{p^{\tau}_i}^{\prime}) \\
        &=\sum_{\eta\in K^{\tau}}d(p^{\tau}_{\eta},p^{\tau}_{j=max(s), s\in \{s_0, s_1, ..., s_{l-1}\}, j\leq \eta})+\sum_{\eta\in K^{\tau}}\Omega(p^{\tau}_{\eta})\\
        &=\sum_{i=0}^{n-1}d(p^{\tau}_i,p^{\tau}_{j=max(s), s\in \{s_0, s_1, ..., s_{l-1}\}, j\leq i})+\Omega(K^{\tau}_{\eta})
    \label{e3}
\end{split}
\end{equation}
\end{small}

Hence, the total weighted distortion over all tiles incurred by $\bm K=\{K^0,K^1,...,K^{t-1}\}$ can be calculated with
\begin{small}
\begin{equation}
\begin{split}
    &\widetilde{D}(\bm K)=\sum_{\tau=0}^{{t-1}}\lambda_{\tau}D(K^{\tau})=\sum_{\tau=0}^{{t-1}}\sum_{\eta \in K^{\tau}}\lambda_{\tau}D(p_{\eta}^{\tau}) \\
    &=\sum_{\tau=0}^{{t-1}}\sum_{\eta \in K^{\tau}}\Big\{\sum_{i=0}^{n-1}\lambda_{\tau}d \big (p^{\tau}_{i},p^{\tau}_{j=max(s), s\in \{s_0, s_1, ..., s_{l-1}\}, j\leq i}\big) \\ 
    &+ \lambda_{\tau}\Omega(p_{\eta}^{\tau})\Big\} \\
    &=\sum_{\tau=0}^{{t-1}} \sum_{\eta \in K^{\tau}} \Big \{\sum_{i=0}^{n-1} \widetilde{d}\big(p^{\tau}_{i},{p^{\tau}_{j=max(s), s\in \{s_0, s_1, ..., s_{l-1}\}, j\leq i}} \big) \\ 
    &+ \widetilde{\Omega}(p_{\eta}^{\tau})\Big\} \\
    &= \sum_{\tau=0}^{{t-1}}\sum_{\eta\in K^{\tau}}\widetilde{D}(P_{\eta}^{\tau})
    \label{e4}
\end{split}
\end{equation}
\end{small}
\noindent where $\lambda_{\tau}$ is the importance factor for the tile $\tau$. We denote the weighted distortion of $\cdot$ as $\widetilde{\cdot}$, e.g., $\lambda_{\tau}D(P_{\eta}^{\tau})$ as $\widetilde{D}(P_{\eta}^{\tau})$.

Let $\omega=n \cdot t$, and the VR video has $\omega$ packets. The packet set $P$ can be written as $P=\{p_0, p_1,...,p_{\omega-1}\}$. Similarly, by assuming that the total number of dropped packets is $f$, the dropped set $K$ can be expressed as $K = \{p_{k_0}, p_{k_1}, ..., p_{k_{f-1}}\}$. In the meantime, the transmission set $S$ with length $g=\omega-f$ can be defined as $S = \{p_{s_0}, p_{s_1}, ..., p_{s_{g-1}}\}$, and $P=K \cup S$. Therefore, by referring to Eq. \ref{e3}, Eq. \ref{e4} can be rewritten as 
\begin{equation}
\begin{split}
    &\widetilde{D}(K)=\sum_{i=0}^{{f-1}}\widetilde{D}(p_{k_i}) \\
    &=\sum_{i=0}^{w-1}\widetilde{d}\big(p_i,p_{j=max(s), s\in \{s_0, s_1, ..., s_{g-1}\}, j\leq i}\big)+\widetilde{\Omega}(K)
    \label{e5} 
\end{split}
\end{equation}
Given available bandwidth, $R_{max}$, the problem can be defined as
\begin{equation}
    K^{*} = \argmin_{K} \widetilde{D}(K), \, s_.t_. \, R(S) \leq R_{max}
    \label{e6}
\end{equation}
which means the system needs to decide on a dropped set $K$ (or a transmission set $S$) to meet the constrained bandwidth causing the minimum weighted distortion. We denote $K=P \backslash S$, where ``$\backslash$'' means the ``set difference'', and $P \backslash S$ is a set consisting of the elements of $P$ which are not elements of $S$. The problem can be rewritten as
\begin{equation}
\begin{split}
    S^{*} &=  \argmin_{S} \widetilde{D}(P \backslash S), s_.t_. R(S) \leq R_{max} \\
    &= \argmin_{S} \Big \{\sum_{i=0}^{\omega -1}\widetilde{d}\big(p_i,p_{j=max(s), s\in \{s_0, s_1, ..., s_{g-1}\}, j\leq i}\big) \\
    &+\widetilde{\Omega}(P \backslash S)\Big\}
    \label{e7}
\end{split}
\end{equation}

\section{Dynamic Programming Solution}
We define $\widetilde{D}({P \backslash S^{s_0:s_{m-1}}_m})$ as the minimum weighted distortion caused by the selected transmission subset $S^{s_0:s_{m-1}}_m = \{p_{s_0}, p_{s_1}, ..., p_{s_{m-1}}\}$ that has m packets, starting with packet $p_{s_0}$ and ending with packet $p_{s_{m-1}}$. According to Eq. \ref{e7}, we have
\begin{small}
\begin{equation}
\begin{split}
    \widetilde{D}({P \backslash S^{s_0:s_{m-1}}_m}) &= \min_{ S^{s_0:s_{m-1}}_m}\Big\{\sum_{i=0}^{\omega -1}\widetilde{d}\big(p_i,p_{j=max(s), s\in S^{s_0:s_{m-1}}_m, j \leq i}\big) \\
    &+\widetilde{\Omega}{(P \backslash S^{s_0:s_{m-1}}_m})\Big\}\\
    \label{e10}
\end{split}
\end{equation}
\end{small}
As the first packet $p_{0}$ is very important (I frame), it is always selected in set $S$ and so we have $s_0=0$. Let $e=s_{m-1}$. Therefore, $s_0=0$ and $s_{m-1}=e$ are removed from the optimization. Hence, Eq. \ref{e10} can be rewritten as
\begin{small}
\begin{equation}
\begin{aligned}
    \widetilde{D}({P \backslash S^{s_0:s_{m-1}}_m})&=\min_{S^{s_1:s_{m-2}}_m}\Big\{\sum_{i=0}^{\omega -1} \widetilde{d}\big(p_i,p_{j=max(s), s\in S^{0:{s_{m-1}}}_{m}, j \leq i}\big) \\
    &+\widetilde{\Omega}({P \backslash S^{s_0:s_{m-1}}_{m}})\Big\}
    \label{e11}
\end{aligned}
\end{equation}
\end{small}
\noindent Similarly, we can define $\widetilde{D}({P \backslash S^{s_0:s_{m-2}}_{m-1}})$ as
\begin{small}
\begin{equation}
\begin{split}
    \widetilde{D}({P \backslash S^{s_0:s_{m-2}}_{m-1}}) &= \min_{ S^{s_1:s_{m-3}}_{m-1}}\Big\{\sum_{i=0}^{w-1}\widetilde{d}\big(p_i,p_{j=max(s), s\in S^{0:s_{m-2}}_{m-1}, j \leq i}\big) \\
    &+\widetilde{\Omega}({P \backslash S^{s_0:s_{m-2}}_{m-1}})\Big\}\\
    \label{e17}
\end{split}
\end{equation}
\end{small}

\noindent Since $0<s_1<s_2<...<s_{m-2}<e$ and $j \leq i$, the Eq. \ref{e11} can be expressed as 
\begin{small}
\begin{equation}
\begin{aligned}
    \widetilde{D}({P \backslash S^{s_0:s_{m-1}}_m})&=\min_{S^{s_1:s_{m-2}}_m}\Big\{\sum_{i=0}^{e-1} \widetilde{d}\big(p_i,p_{j=max(s), s\in S^{0:s_{m-2}}_{m-1} , j \leq i}\big)\\
    & +\sum_{i=e}^{\omega -1}\widetilde{d}\big(p_i,p_{j=max(s), s\in S^{0:s_{m-2}}_{m-1}, j \leq i}\big) \\
    & -\sum_{i=e}^{\omega -1}\widetilde{d}\big(p_i,p_{j=max(s), s\in S^{0:s_{m-2}}_{m-1}, j \leq i}\big) \\
    & +\widetilde{\Omega}({P \backslash S^{s_0:s_{m-1}}_{m}})\Big\}\\
    & +\sum_{i=e}^{\omega -1} \widetilde{d}(p_i, p_e)
    \label{e13}
\end{aligned}
\end{equation}
\end{small}
\noindent Because of $s_{m-2}<e$, we have
\begin{equation}
\begin{aligned}
\sum_{i=e}^{\omega -1} \widetilde{d}\big(p_i,p_{j=max(s), s\in S^{0:s_{m-2}}_{m-1}, j \leq i})=\sum_{i=e}^{\omega -1}\widetilde{d}(p_i, p_{s_{m-2}}\big)
\end{aligned}    
\end{equation}
Therefore, Eq. \ref{e13} can be rewritten as
\begin{small}
\begin{equation}
\begin{split}
\widetilde{D}({P \backslash S^{s_0:s_{m-1}}_m})
&=\min_{S^{s_1:s_{m-2}}_m}\Big\{\sum_{i=0}^{\omega -1} \widetilde{d}\big(p_i,p_{j=max(s), s\in S^{0:s_{m-2}}_{m-1}, j \leq i}\big) \\
&+\widetilde{\Omega}({P \backslash S^{s_0:s_{m-1}}_m})\\
&-\sum_{i=e}^{\omega -1} \big(\widetilde{d}(p_i, p_{s_{m-2}})-\widetilde{d}(p_i, p_e)\big) \Big\}
\label{e14}
\end{split}
\end{equation}
\end{small}

\noindent Since $\widetilde{\Omega}({P \backslash S^{s_0:s_{m-1}}_m})=\widetilde{\Omega}({P \backslash S^{s_0:s_{m-2}}_{m-1}})-\widetilde{\Omega}(e)$, Eq. \ref{e14} can be expressed as 
\begin{small}
\begin{equation}
\begin{split}
\label{e111}
\widetilde{D}({P \backslash S^{s_0:s_{m-1}}_m}) &=\min_{S^{s_1:s_{m-2}}_m}\Big\{\sum_{i=0}^{n-1}\widetilde{D}\big(p_i,p_{j=max(s), s\in S^{0:s_{m-2}}_{m-1}, j \leq i}\big) \\
&+\widetilde{\Omega}({P \backslash S^{s_0:s_{m-2}}_{m-1}})-\widetilde{\Omega}(e)\\
&-\sum_{i=e}^{\omega -1} \big(\widetilde{d}(p_i, p_{s_{m-2}})-\widetilde{d}(p_i, p_e)\big) \Big\}
\end{split}    
\end{equation}
\end{small}

\noindent Let 
\begin{equation}
    \Phi(s_{m-2}, e)= \widetilde{\Omega}(e) + \sum_{i=e}^{\omega -1} \big(\widetilde{d}(p_i, p_{s_{m-2}})-\widetilde{d}(p_i, p_e)\big)
\end{equation}
Eq. \ref{e111} can be rewritten as
\begin{equation}
\begin{split}
    \widetilde{D}({P \backslash S^{s_0:s_{m-1}}_m}) 
    &=\min_{S^{s_1:s_{m-2}}_m}\Big \{\sum_{i=0}^{n-1} \widetilde{d}\big(p_i,p_{j=max(s), s\in S^{0:s_{m-2}}_{m-1}, j \leq i}\big) \\
    &+\widetilde{\Omega}({P \backslash S^{s_0:s_{m-2}}_{m-1}}) -\Phi(s_{m-2}, e) \Big\}
    \label{e18}
\end{split}
\end{equation}
By referring to Eq. \ref{e17}, the Eq. \ref{e18} can be rewritten as
\begin{equation}
\begin{split}
     &\widetilde{D}({P \backslash S^{s_0:s_{m-1}}_m})= \\
     & \min_{s_{m-2}}\bigg\{\min_{S^{s_1:s_{m-3}}_{m-1}}\Big\{\sum_{i=0}^{\omega -1} \widetilde{d}\big(p_i,p_{j=max(s), s\in S^{0:s_{m-2}}_{m-1}, j \leq i }\big) \Big\} \\
    &+\widetilde{\Omega}({P \backslash S^{s_0:s_{m-2}}_{m-1}}) - \Phi(s_{m-2}, e)\bigg\} \\
    &=\min_{s_{m-2}}\Big\{\widetilde{D}({P \backslash S^{s_0:s_{m-2}}_{m-1}})-\Phi(s_{m-2}, e)\Big\}
    \label{24}
\end{split}
\end{equation}
where the first part represents the minimized distortion for the set $S^{s_0:s_{m-2}}_{m-1}$, while the second part is the distortion reduction if packet $e$ is selected into the set $S^{s_0:s_{m-2}}_{m-1}$ to generate the new set $S^{s_0:s_{m-1}}_{m}$. Eq. \ref{24} describes the state transition in dynamic programming. As the first frame is always selected into $S$, the initial state $S^0_{1}$ can be defined as
\begin{equation}
    D(P\backslash S^0_{1})=\sum_{e=1}^{m-1}d(p_0,p_e)+\Omega(P \backslash S^0_{1})
\end{equation}

With the initial state and state transition equation, the system can compute the optimal packet scheduling scheme via backtracking trellis diagram. As shown in the trellis diagram of Fig.\ref{view}, the trellis node represents a transmission set that has a corresponding edge cost (distortion). For instance, the transmission set $S^0_1$ indicates the current set has 1 packet and ends with packet 0 as the packet 0 is always selected. $\Phi(S_1^0,4)$ indicates the distortion cost when packet 4 is selected into set $S^0_1$ and generates a new set $S^4_2$. Meanwhile, the remaining bandwidth for the transmission set can be computed. If the value is 0, the transmission set is the final set leading to maximum distortion.

Suppose there are n packets in a network node, the computational complexity of the proposed solution is $O(n^2)$. Since a frame is decomposed into several (m) packets in the real world, the computational complexity is $O({(\frac{n}{m})}^2)$.
\section{Performance of Viewpoint Prediction Model}
\subsection{Implementation Details}
The dataset used in our work is MMSys18 \cite{10.1145/3204949.3208139}, which contains 1083 viewpoint trajectories obtained from 57 participants with 19 4K videos. The dataset is randomly split into training and testing sets, where the training set contains trajectories from the intersection of 70\% of videos (13 videos) and 50\% of users (24 users) by following the configuration in Track \cite{9395242}. The videos are split into 24 tiles. The viewport is defined as a 120$\degree$ $\times$ 120$\degree$ area. The time window of the input is set to 1 second. The model is tested with different output time windows from 1 to 5 seconds. Nonetheless, our system uses the 1-second future output. Five points and frames are sampled every second. Therefore, the model input has a sequence of 5 points and a sequence of 5 frames. The visual encoder has 4 temporal and spatial encoder blocks while the viewpoint encoder has 2 transformer encoder blocks. The viewpoint decoder has 2 decoder blocks. The model uses 12 multi-head attention and an embedding dimension of 768. The batch size is 100. The learning rate is 0.0005 with a decay rate of 0.99. 

\subsection{Results and Analysis}
In our experiments, average great-circle distance is used as the evaluation metric, which is the shortest distance between the ground truth point and the predicted point on the surface of a sphere. The proposed method is compared to three methods on the dataset MMSys18, as shown in Table \ref{tf_results}. The baseline method uses the last input element as the output trajectory. The VPT360 \cite{9733647} only uses the encoder of the transformer. The Track \cite{9395242} is based on the LSTM. 
The results of VPT360 and Track are obtained from the paper and the pre-trained model, respectively.

\begin{table}[htbp]
    \caption{Comparison of viewpoint predictions for average great circle distance on MMSys18.}
    \centering
    \resizebox{0.45\textwidth}{1.2cm}{\begin{threeparttable}
    \begin{tabular}{ |c| c| c|c|c|c| }
    \hline
\multirow{2}{*}{Method}                        & \multicolumn{5}{c|}{Prediction Time Window}                                                                                                                                                                                                                                \\ \cline{2-6} 
                                               & \multicolumn{1}{c|}{1st s\tnote{$\ast$}}                              & \multicolumn{1}{c|}{2nd s}                              & \multicolumn{1}{c|}{3rd s}                              & \multicolumn{1}{c|}{4th s}                              & 5th s                              \\ \hline
     Baseline & 0.321 & 0.520 & 0.672 & 0.788 & 0.878\\  
     \hline
     VPT360 \cite{9733647} & 0.215 & 0.399 & 0.550 & 0.667 & 0.754 \\
     \hline
     Track \cite{9395242} & 0.281 & 0.464 & 0.609 & 0.723 & 0.809 \\
     \hline
     Ours & \textbf{0.212} & \textbf{0.386} & \textbf{0.531} & \textbf{0.643} & \textbf{0.723} \\
     \hline
    \end{tabular}
    
    \begin{tablenotes}
        \footnotesize               
            \item[$\ast$] The s indicates second.  
    \end{tablenotes}
    \end{threeparttable}}
    \label{tf_results}
\end{table}

The bold numbers are the best scores. It can be seen that the average distances of all methods increase as the prediction time window increases. The proposed transformer performs best compared to other methods across different time windows.

\section{Evaluation of Packet Scheduling Strategy}

\subsection{Implementation Details}
Four existing methods are compared to our method: baseline which is the tail drop algorithm ignoring the difference between packets, EWRD (equal-weighted rate-distortion) \cite{8836788} dropping packets based on real distortion by considering tiles equally, NIRAP \cite{https://doi.org/10.48550/arxiv.2202.04703} preemptively dropping non-IRAP packets and IPB \cite{MOHARRAMI2023100149} dropping packets based on frame types. The performances of all methods are evaluated on five metrics: total distortion, viewport distortion, total packet loss, viewport packet loss, and viewport bandwidth consumption.
All videos (6 videos) of the testing set from MMSys18 are evaluated.
All methods are evaluated on two network trace scenarios: constant bandwidth trace (0.5-30 Mbps) and real-world trace. A 4G LTE dataset with throughput ranging from 0 to 173 Mbps \cite{10.1145/3204949.3208123} is used for real-world testing. The mean and STD (standard deviation) are presented in both scenarios.

\begin{figure*}[htbp]
    \centering
    \subfloat[]{
        \includegraphics[width=0.48\textwidth]{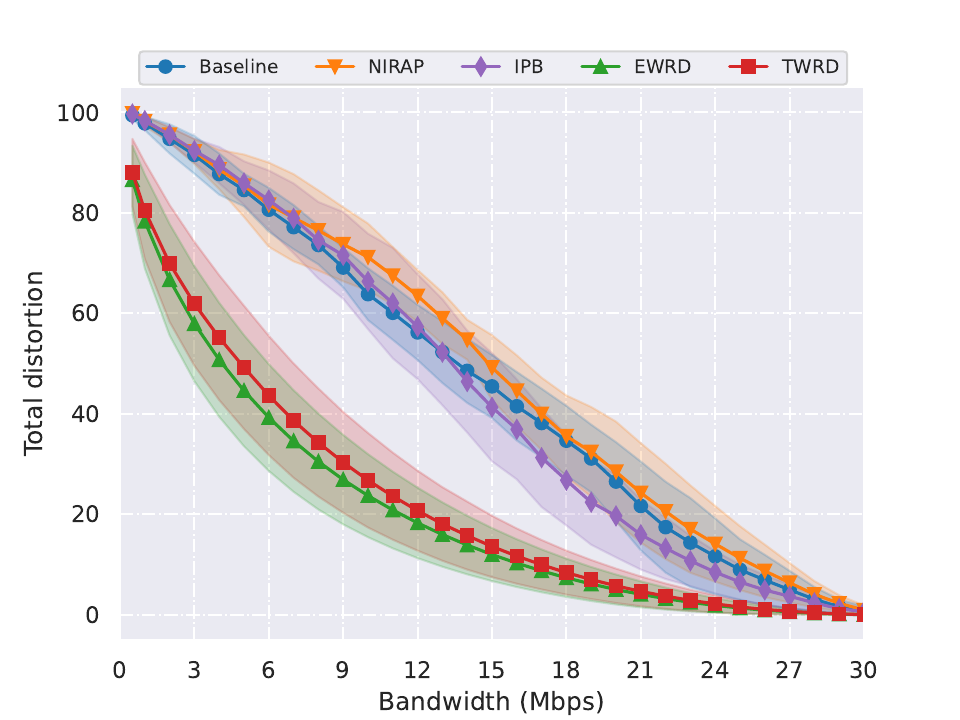}
    }
    \subfloat[]{
	\includegraphics[width=0.48\textwidth]{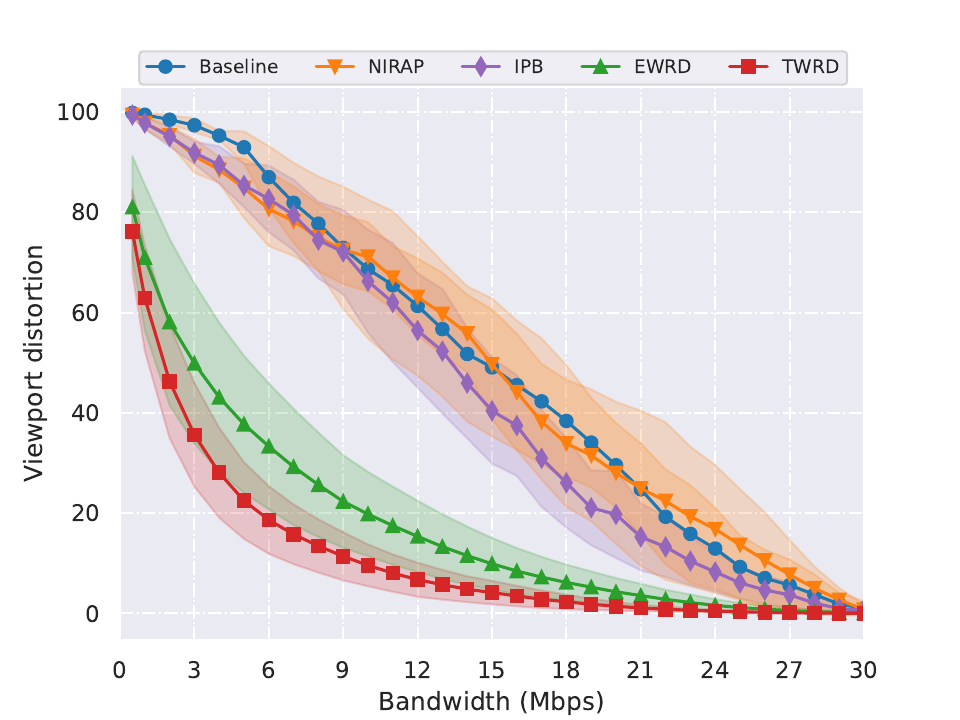}
    }
        \hfill
    \subfloat[]{
        \includegraphics[width=0.48\textwidth]{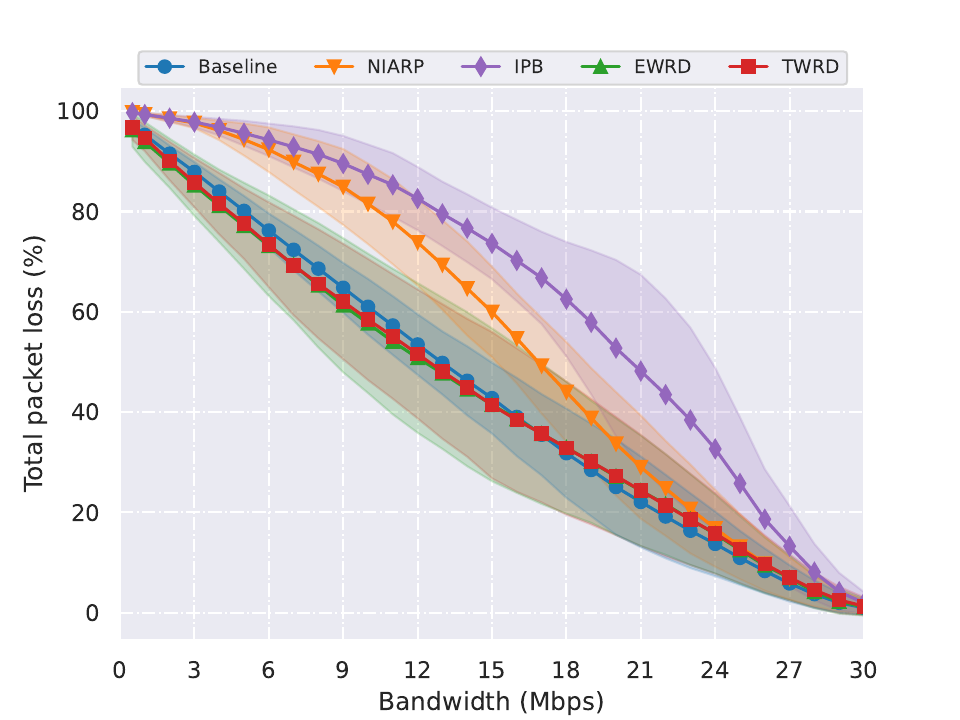}
    }
    \subfloat[]{
    \includegraphics[width=0.48\textwidth]{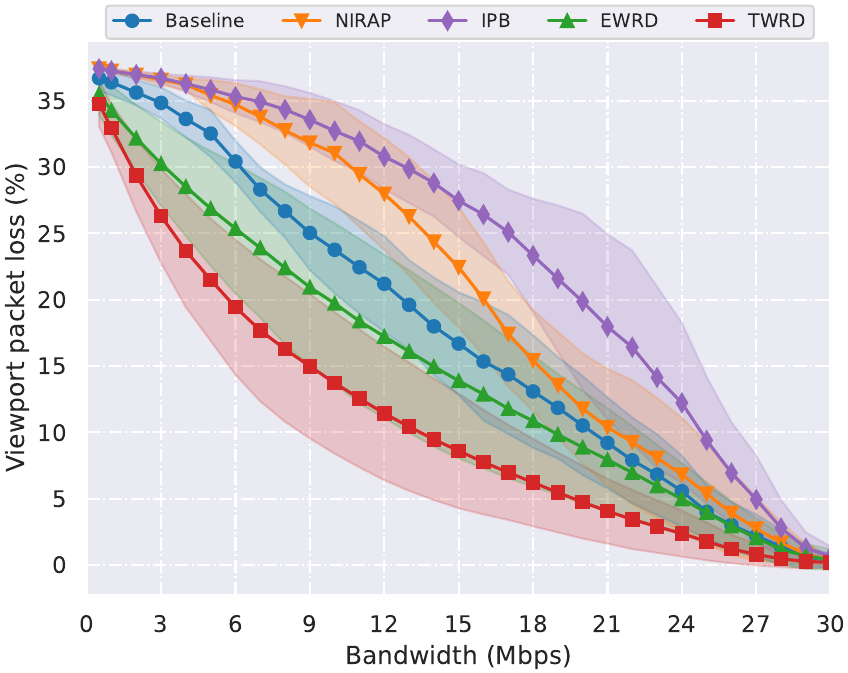}
    }
        \hfill
    \subfloat[]{
	\includegraphics[width=0.48\textwidth]{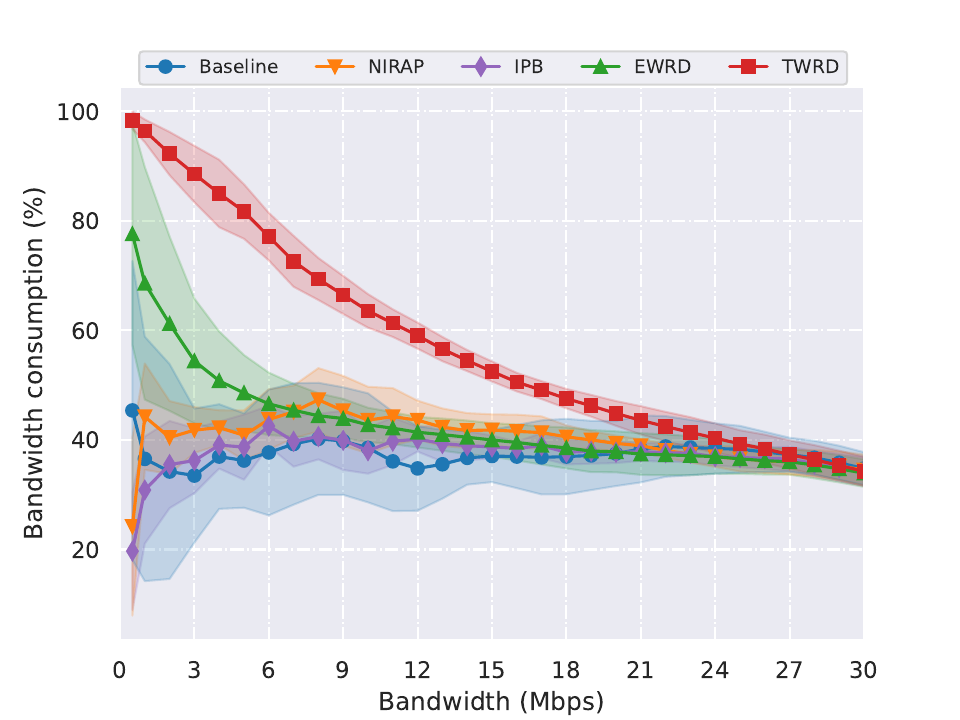}
    }
    \subfloat[]{
	\includegraphics[width=0.48\textwidth]{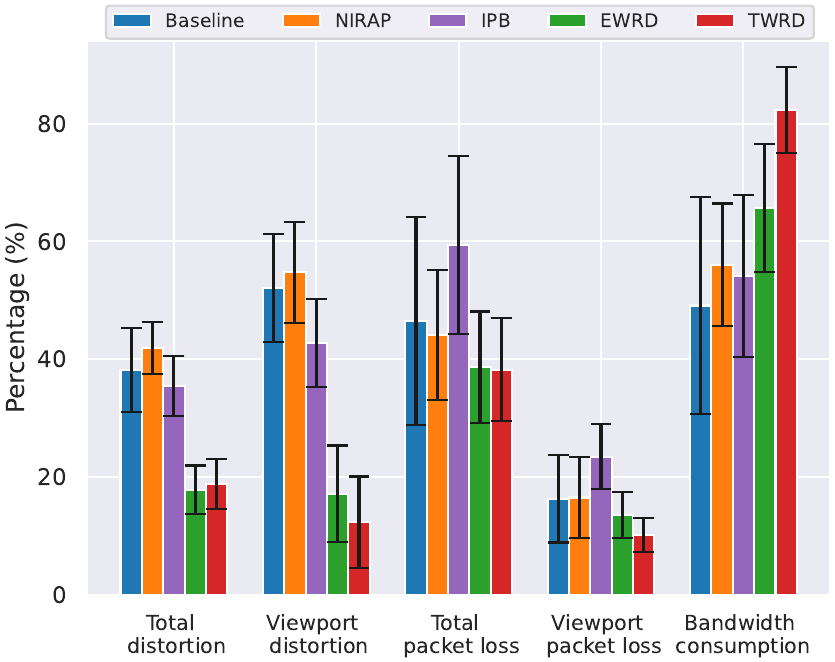}
    }
    \caption{Experiment results for the five methods under different bandwidth scenarios. Constant bandwidth scenario: (a) Total distortion. (b) Viewport distortion. (c) Total packet loss. (d) Viewport packet loss. (e) Bandwidth consumption.  Real-world trace scenario: (f) Experiment results on the five metrics under the 4G LTE dataset. }
     \label{dis}
\end{figure*}

\subsection{Constant Bandwidth Scenario}
\subsubsection{Distortion Analysis}
The total distortions with different packet scheduling strategies across various bandwidths are shown in Fig. \ref{dis}(a). Distortion values are normalized and scaled to a range of 0 to 100 for comparison. The baseline, NIRAP and IPB have similar distortion while the TWRD and EWRD perform better and have similar distortion. There is not much difference between the TWRD and EWRD, while the EWRD outperforms the others. The difference comes from the fact that EWRD uses knowledge about how dropping packets affect reconstructed video quality, while TWRD relies on weighted distortion. Thus, EWRD drops packets that have the least effect on reconstructed videos. Meanwhile, the distortions of all methods decrease with the bandwidth increasing.

The TWRD performs best on viewport distortion throughout the range, as shown in Fig. \ref{dis}(b), and the EWRD achieves the second-best performance, followed by the baseline, IPB, and NIRAP. For example, at the bandwidth of 0.5 Mbps, the improvements of TWRD are 28.5\%, 28.4\%, 27.8\%, and 6.4\%, over the baseline, NIRAP, IPB, and EWRD, respectively. As bandwidth varies, the improvement varies. The TWRD exhibits average enhancements of approximately 74\%, 76\%, 70\%, and 35\%, when compared to the baseline, NIRAP, IPB, and EWRD, respectively, across all bandwidths. Meanwhile, the average STDs across all bandwidths are 10.5, 13.3, 10.2, 7.6, and 4.5 for baseline, NIRAP, IPB, EWRD, and TWRD, respectively. Therefore, the quality of the viewport is greatly improved with the proposed approach compared to the other methods.

\renewcommand{\multirowsetup}{\centering}
\begin{table*}[htpb]
\centering
\caption{Case analysis for the five methods on the two videos: PortoRiverside and Warship.} 
\label{two}
\resizebox{0.99\textwidth}{4.3cm}{
\begin{threeparttable}
\begin{tabular}{|c|c|c|c|c|c|c|c|c|c|c|c|c|c|c|}
\hline
Video & \multicolumn{7}{|c|}{PortoRiverside} & \multicolumn{7}{|c|}{Warship} \\ \hline

        Bandwidth        & 0.5 Mbps         & 5 Mbps       & 10 Mbps       & 15 Mbps      & 20 Mbps      & 25 Mbps       & 30 Mbps        & 0.5 Mbps      & 5 Mbps        & 10 Mbps       & 15 Mbps     & 20 Mbps       & 25 Mbps       & 30 Mbps     \\ \hline

                                 \multirow{5}{*}{\begin{tabular}[c]{clc}Total\\ distortion\end{tabular}}  & 941.0 & 730.9 & 617.5           & 404.4           & 191.6           & 27.0            & 0     & 1143.9 & 1021.0 &  637.4           & 501.6           & 259.3           & 119.0           & 12.6           \\ \cline{2-15}
                                                                                                             & 954.6 & 800.4 & 722.5           & 546.0           & 379.3           & 81.2            & 0     & 1144.2 & 1019.7 & 883.8           & 714.1           & 440.2           & 149.5           & 19.0          \\ \cline{2-15}
                                                                                                              & 951.5 & 824.8 & 659.0           & 505.3           & 238.6           & 33.2            & 0        & 1142.9 & 1037.5 & 854.3           & 564.3           & 347.8           & 106.0           & 11.9       \\ \cline{2-15} 
                                                                                                             & \textbf{786.2} & \textbf{383.6} & \textbf{209.2}  & \textbf{112.9}  & \textbf{44.0}   & \textbf{5.6}    & 0     & \textbf{940.2} & \textbf{529.0} &\textbf{322.4}  & \textbf{185.4}  & \textbf{93.0}   & \textbf{31.7}   & \textbf{2.9}        \\ \cline{2-15} 
                                                                                                             & 800.9 & 422.1 & 233.1           & 125.3           & 51.9            & 6.7             & 0      & 942.4 & 563.8 &357.2           & 207.4           & 106.6           & 38.0            & 3.6         \\ \cline{1-15}

                                  \multirow{5}{*}{\begin{tabular}[c]{clc}Viewport\\ distortion\end{tabular}}  
                                                                                                            & 447.7 & 342.3 & 170.3           & 142.0           & 115.0            & 30.6            & 0    & 1273.2 & 1202.7 & 268.5           & 235.8           & 69.0            & 19.3            & 1.7           \\ \cline{2-15}
                                                                                                             & 454.8 & 369.5 & 312.4        & 229.5           & 67.1            & 6.2            & 0    & 635.2 & 553.6 & 492.9           & 432.9           & 243.5            & 27.8            & 2.4            \\ \cline{2-15}
                                                                                                             & 455.2 & 403.8 & 345.5        & 233.6           & 117.7            & 18.4            & 0   & 635.4 & 583.8 & 469.3           & 300.9           & 208.8            & 49.6            & 4.5            \\ \cline{2-15}
                                                                                                             & 328.7 & 125.0 & 66.4         & 31.9            & 17.3            & 2.8             & 0    & {429.8} & 165.6 & 96.2            & 59.5            & 31.4            & 13.9            & 1.6             \\ \cline{2-15}
                                                                                                             & \textbf{313.8} & \textbf{75.1} &\textbf{28.9}   & \textbf{10.3}   & \textbf{4.4}    & \textbf{0.6}    & 0     & \textbf{429.8} & \textbf{105.4} &\textbf{43.3}   & \textbf{23.0}   & \textbf{9.8}    & \textbf{3.0}    & \textbf{0.3}          \\ \cline{1-15}
                                 
                                  \multirow{5}{*}{\begin{tabular}[c]{clc}Total\\ packet loss\end{tabular}}     
                                                                                                             & \textbf{95.5}\% & 76.7\% & 59.7\%          & 39.7\%          & 22.9\%          & 6.6\%           & 0\%       & 99.1\% & 98.3\% &   63.4\%          & \textbf{45.6}\% & \textbf{29.6\%} & \textbf{13.6\%} & 1.5\%       \\ \cline{2-15}
                                                                                                              & 99.9\% & 96.7\% &92.4\%          & 76.2\%          & 38.1\%          & 7.5\%           & 0\%     & 99.8\% & 96.2\% & 86.2\%          & 61.5\% & 38.6\% & 17.9\% & 2.1\%         \\ \cline{2-15}
                                                                                                              & 99.9\% & 97.9\% &95.1\%          & 85.5\%          & 66.6\%          & 24.0\%           & 0\%       & 99.7\% & 96.9\% & 91.4\%          & 77.6\% & 67.1\% & 41.2\% & 4.9\%       \\ \cline{2-15}
                                                                                                              & 96.0\% & \textbf{65.8}\% & \textbf{41.1\%} & \textbf{29.5\%} & 20.7\%          & \textbf{5.8\%}  & 0\%  & \textbf{98.1}\% & 82.1\% & 65.4\%          & 49.5\%          & 34.2\%          & 17.1\%         & \textbf{1.4\%}           \\ \cline{2-15}
                                                                                                              & 95.7\% & 68.6\% &45.2\%          & 29.7\%          & \textbf{18.9\%} & 6.2\%           & 0\%    & 98.2\% & \textbf{79.7}\% &\textbf{63.4\%} & 48.8\%          & 34.5\%          & 18.9\%          & 2.6\%          \\ \cline{1-15} 
                                                                                              
                                  \multirow{5}{*}{\begin{tabular}[c]{clc}Viewport\\ packet loss\end{tabular}}   
                                                                                                             & 35.1\% & 25.0\% &  25.1\%          & 17.2\%          & 7.2\%           & 3.0\%           & 0\%    & 36.6\% & 35.4\% &   20.2\%          & 15.8\%          & 8.5\%           & 2.0\%           & 0.1\%          \\ \cline{2-15} 
                                                                                                              & 37.5\% & 36.0\% & 35.0\%          & 30.9\%          & 19.5\%           & 4.3\%           & 0\%   & 37.3\% & 36.1\% & 33.3\%          & 23.1\%          & 11.2\%           & 2.2\%           & 0.2\%           \\ \cline{2-15} 
                                                                                                              & 37.5\% & 36.7\% & 35.8\%          & 32.5\%          & 25.0\%           & 8.9\%           & 0\%    & 37.4\% & 36.3\%  & 34.2\%          & 29.2\%          & 25.5\%           & 15.6\%           & 1.6\%          \\ \cline{2-15} 
                                                                                                              & 35.5\% & 22.8\% & 14.3\%          & 10.1\%          & 7.4\%           & 2.0\%           & 0\%     & \textbf{35.8}\% & 26.0\% & 19.7\%          & 14.2\%          & 9.3\%           & 4.3\%           & 0.2\%     \\ \cline{2-15} 
                                                                                                              & \textbf{33.7}\% & \textbf{16.9}\% &\textbf{9.6\%}  & \textbf{6.0\%}  & \textbf{3.6\%}  & \textbf{1.0\%}  & 0\%     & \textbf{35.8}\% & \textbf{19.6}\% &\textbf{11.4\%} & \textbf{7.2\%}  & \textbf{3.2\%}  & \textbf{0.8\%}  & \textbf{0\%}        \\ \cline{1-15} 
                                  \multirow{5}{*}{\begin{tabular}[c]{clc}Bandwidth\\ consumption \end{tabular}}
                                                                                                             & 49.3\% & 47.5\% & 30.0\%          & 30.3\%          & 37.1\%          & 33.8\%          & 29.7\%     & 70.6\% & 33.5\% &  42.9\%          & 33.5\%          & 33.7\%          & 34.7\%          & 30.1\%      \\ \cline{2-15}
                                                                                                              & 17.5\% & 46.7\% & 28.2\%          & 34.0\%          & 31.4\%          & 33.1\%          & 29.7\%    & 11.2\% & 43.5\% & 43.2\%          & 39.5\%          & 40.7\%          & 40.8\%          & 35.7\%      \\ \cline{2-15}
                                                                                                              & 13.4\% & 29.3\% & 35.2\%          & 35.0\%          & 35.5\%          & 34.1\%          & 29.7\%    & 16.7\% & 44.9\% & 46.4\%          & 41.7\%          & 38.7\%          & 38.1\%          & 35.4\%        \\ \cline{2-15}
                                                                                                              & 73.1\% & 47.7\% & 40.6\%          & 38.3\%          & 33.7\%          & 33.6\%          & 29.7\%   & 96.7\% & 59.3\% & 46.1\%          & 39.5\%          & 36.9\%          & 35.1\%          & 34.8\%        \\ \cline{2-15}
                                                                                                              & \textbf{100.0}\% & \textbf{78.2}\% &\textbf{61.6\%} & \textbf{51.0\%} & \textbf{41.4}\% & \textbf{35.1}\% & 29.7\%  & \textbf{99.5}\% & \textbf{89.1}\% &\textbf{68.2\%} & \textbf{53.9\%} & \textbf{46.4\%} & \textbf{40.8\%} & \textbf{35.7\%}        \\ \hline 
                                          
\end{tabular}%
    \end{threeparttable}}
    \begin{tablenotes}
        \footnotesize              
        \item[1] Note: The values for the five methods are listed in the order: baseline, NIRAP, IPB, EWRD, and TWRD for every metric.  
    \end{tablenotes}

\end{table*}

\subsubsection{Packet Loss Analysis}

The total and viewport packet loss rates for all strategies are shown in Fig. \ref{dis}(c) and (d). The total packet loss rate denotes the percentage of lost packets relative to the total transmitted packets. Simultaneously, the viewport packet loss rate indicates the percentage of lost packets in the viewport area compared to the total packets, which degrades video quality. Given that the viewport consists of a maximum of 9 tiles, representing 37.5\% of all tiles, the highest possible viewport packet loss rate is 37.5\%. The total and viewport packet loss rates decrease as the bandwidth increases. Since packets have different distortions, the loss of a small number of packets can cause the same or greater distortion than the loss of a large number of packets. Among the various methods, IPB exhibits the highest overall packet loss rate. As the B-type packets usually have small sizes, more B-type packets are dropped to meet the bandwidth constraint. Although the total packet loss of the NIRAP is lower than the IPB, the NIRAP drops P-type and B-type packets, which causes the highest total distortion. TWRD avoids dropping packets in the viewport, which sometimes causes a higher total packet loss rate. 

Although the baseline, EWRD, and TWRD have similar total packet loss rates across all bandwidths, their viewport packet loss rates present huge differences, as shown in Fig. \ref{dis}(d). Noticeably, our method still performs well in terms of viewport packet loss, even under severely limited bandwidth conditions, such as 0.5 Mbps, where a majority of packets are dropped. The improvements of TWRD over EWRD, IPB, NIRAP, and the baseline are approximately 2.1\%, 7.1\%, 7.1\%, and 4.8\%, respectively, at the bandwidth of 0.5 Mbps. As viewport packets are weighted and rarely dropped, TWRD always has the lowest viewport packet loss across bandwidths.

\subsubsection{Bandwidth Consumption Analysis}

Since the simulated system has limited bandwidth without background traffic, the total bandwidth consumption for the entire VR video is the bandwidth itself. The percentage of viewport bandwidth consumption indicates how much bandwidth resource is allocated to the viewport. The higher the value is, the better the method is. The viewport bandwidth consumption is visualized, as shown in Fig. \ref{dis}(e). The TWRD consumes the most bandwidth, followed by the EWRD. The baseline, NIRAP, and IPB have similar bandwidth consumption rates. When the bandwidth is very limited, for instance, 0.5 Mbps, most packets including packets inside the viewport are dropped. Therefore, the baseline, NIRAP, and IPB have few resources allocated for the viewport while TWRD prioritizes saving packets within the viewport, with TWRD achieving nearly 100\% bandwidth utilization. Because the bandwidth resource required by the viewport is constant, the percentages of the TWRD and EWRD drop down gradually with bandwidth increase. Noticeably, the values of the baseline, NIRAP, and IPB fluctuate at about 40\% over bandwidths, as packets are dropped randomly, and every tile has the same probability of being dropped. The difference between all methods gradually decreases with bandwidth until sufficient bandwidth is provided.

\subsection{Real-world Trace Scenario}
The means and STDs across all videos and network traces are listed in Fig. \ref{dis}(f). The TWRD performs best on four metrics except for total distortion, followed by EWRD, while the baseline, NIRAP, and IPB have poor performance. For example, the baseline, NIRAP, IPB, EWRD, and TWRD exhibit viewport distortions of 0.52, 0.55, 0.43, 0.17, and 0.12, respectively, accompanied by STDs of 9.2, 8.5, 7.5, 8.2, and 7.8. Additionally, the viewport packet loss rates for them are 16.3, 16.5, 23.5, 13.6, and 10.21 with STDs of 7.4, 6.9, 5.5, 3.9, and 2.9, respectively. TWRD's advantages and robustness are demonstrated by the experiment results on constant bandwidths and real-world traces.

\subsection{Case Study Analysis}
We also randomly choose two videos, Warship and PortoRiverside, and provide the detailed values for both videos in Table \ref{two} to provide an in-depth analysis.
The two videos are analyzed with the five metrics under different bandwidths (0.5 Mbps, 5 Mbps, 10 Mbps, 15 Mbps, 20 Mbps, 25 Mbps, 30 Mbps). The values are listed in the order baseline, NIRAP, IPB, EWRD, and TWRD for every metric. Both videos present similar trends with different values due to the difference between the videos and the randomness of the baseline, NIRAP, and IPB. For both videos, the EWRD has the smallest overall distortion, and it is close to the values of TWRD. The proposed method still performs better than the others regarding viewport distortion. 
For the viewport packet loss rate, the proposed method has the lowest loss rate across bandwidths, which means TWRD drops more packets that are beyond the viewport. In terms of viewport bandwidth consumption, TWRD always allocates high bandwidth resources to its viewport.

\subsection{Runtime Analysis}
All experiments are conducted using a testbed built on a computer with an Intel Core i9-11900F CPU and an Nvidia GTX3090 GPU. The transformer model has a runtime of around 63.1 ms for the 1-second predicted window, indicating the runtime is acceptable. 

Besides the transformer, the dynamic programming solution is activated only if the queue exceeds a threshold, which is 70\% of the buffer size in our work. The average runtime of the dynamic programming is about 78.2 ms. However, since the packet scheduling is activated only if the queue exceeds a threshold, it has a limited impact on the system delay. Meanwhile, the method drops packets in advance to avoid congestion, which can also reduce system latency. The runtime of our system on the real-world trace dataset is about 86.5 ms.

\section{Conclusion}
In this paper, we propose a TWRD packet scheduling system for VR video streaming based on a multimodal spatial-temporal attention transformer. The problem of determining the optimal packet scheduling scheme is modeled as an optimization problem and solved with the proposed dynamic programming-based solution. The experiment results show that the proposed method reduces the viewport distortion and achieves better performance than the other methods.

Furthermore, our approach can be improved in the future. In practice, our packet scheduling method may introduce a time cost during video streaming. Therefore, future work should focus on optimizing temporal cost reduction to ensure efficient and seamless video delivery to users.
Additionally, since the proposed method reduce quality distortion at the network layer, it can be integrated with various application-layer strategies (e.g. rate adaptation, scalable coding or re-encoding) to further enhance user experience. For instance, scalable coding can be adopted to save bandwidth by buffering the base layer for a long period and fetching enhancement layers for a shorter period to increase the quality of viewport \cite{10.1145/3123266.3123414}. 
Simultaneously, frame-level coding and streaming can be implemented, where coding is concentrated on the viewport, and other regions are encoded progressively, facilitating the video's periodic refreshment \cite{10.1145/3394171.3413751}.

\def\refname{REFERENCES}

\bibliographystyle{IEEEtran}
\bibliography{ref.bib}

\begin{thebibliography}{10}
\providecommand{\url}[1]{#1}
\csname url@samestyle\endcsname
\providecommand{\newblock}{\relax}
\providecommand{\bibinfo}[2]{#2}
\providecommand{\BIBentrySTDinterwordspacing}{\spaceskip=0pt\relax}
\providecommand{\BIBentryALTinterwordstretchfactor}{4}
\providecommand{\BIBentryALTinterwordspacing}{\spaceskip=\fontdimen2\font plus
\BIBentryALTinterwordstretchfactor\fontdimen3\font minus \fontdimen4\font\relax}
\providecommand{\BIBforeignlanguage}[2]{{%
\expandafter\ifx\csname l@#1\endcsname\relax
\typeout{** WARNING: IEEEtran.bst: No hyphenation pattern has been}%
\typeout{** loaded for the language `#1'. Using the pattern for}%
\typeout{** the default language instead.}%
\else
\language=\csname l@#1\endcsname
\fi
#2}}
\providecommand{\BIBdecl}{\relax}
\BIBdecl

\bibitem{9927197}
R.~Tan, R.~Gao, W.~Li, K.~Cao, Y.~Li, C.~Lv, F.-Y. Wang, and D.~Cao, ``Xsickness in intelligent mobile spaces and metaverses,'' \emph{IEEE Intelligent Systems}, vol.~37, no.~5, pp. 86--94, 2022.

\bibitem{https://doi.org/10.48550/arxiv.2202.04703}
L.~Gobatto, M.~Saquetti, C.~Diniz, B.~Zatt, W.~Cordeiro, and J.~R. Azambuja, ``Improving content-aware video streaming in congested networks with in-network computing,'' in \emph{Proceedings of IEEE International Symposium on Circuits and Systems}, 2022, pp. 1813--1817.

\bibitem{7574700}
X.~Corbillon, F.~Boyrivent, G.~A. De~Williencourt, G.~Simon, G.~Texier, and J.~Chakareski, ``Efficient lightweight video packet filtering for large-scale video data delivery,'' in \emph{Proceedings of IEEE International Conference on Multimedia \& Expo Workshops}, 2016, pp. 1--6.

\bibitem{6172571}
Y.-L. Chang, T.-L. Lin, and P.~C. Cosman, ``Network-based {H.264/AVC} whole-frame loss visibility model and frame dropping methods,'' \emph{IEEE Transactions on Image Processing}, vol.~21, no.~8, pp. 3353--3363, 2012.

\bibitem{MOHARRAMI2023100149}
A.~Moharrami, M.~Ghasempour, and M.~Ghanbari, ``A smart packet type identification scheme for selective discard of video packets,'' \emph{e-Prime - Advances in Electrical Engineering, Electronics and Energy}, vol.~4, p. 100149, 2023.

\bibitem{1608103}
J.~Chakareski and P.~Frossard, ``Rate-distortion optimized distributed packet scheduling of multiple video streams over shared communication resources,'' \emph{IEEE Transactions on Multimedia}, vol.~8, no.~2, pp. 207--218, 2006.

\bibitem{4543842}
Y.~Li, A.~Markopoulou, J.~Apostolopoulos, and N.~Bambos, ``Content-aware playout and packet scheduling for video streaming over wireless links,'' \emph{IEEE Transactions on Multimedia}, vol.~10, no.~5, pp. 885--895, 2008.

\bibitem{NASRALLA2018126}
M.~M. Nasralla, M.~Razaak, I.~U. Rehman, and M.~G. Martini, ``Content-aware packet scheduling strategy for medical ultrasound videos over {LTE} wireless networks,'' \emph{Computer Networks}, vol. 140, pp. 126--137, 2018.

\bibitem{9068423}
I.-S. Comşa, G.-M. Muntean, and R.~Trestian, ``An innovative machine-learning-based scheduling solution for improving live {UHD} video streaming quality in highly dynamic network environments,'' \emph{IEEE Transactions on Broadcasting}, vol.~67, no.~1, pp. 212--224, 2021.

\bibitem{9086630}
J.~Chakareski, ``Viewport-adaptive scalable multi-user virtual reality mobile-edge streaming,'' \emph{IEEE Transactions on Image Processing}, vol.~29, pp. 6330--6342, 2020.

\bibitem{8836788}
A.~Abdelhadi, A.~Gerstlauer, and S.~Vishwanath, ``Real-time rate distortion optimized and adaptive low complexity algorithms for video streaming,'' in \emph{Proceedings of 2019 IEEE International Systems Conference}, 2019, pp. 1--8.

\bibitem{10.5555/3295222.3295349}
A.~Vaswani, N.~Shazeer, N.~Parmar, J.~Uszkoreit, L.~Jones, A.~N. Gomez, L.~Kaiser, and I.~Polosukhin, ``Attention is all you need,'' in \emph{Proceedings of the 31st International Conference on Neural Information Processing Systems}, 2017, p. 6000–6010.

\bibitem{vit}
A.~Dosovitskiy, L.~Beyer, A.~Kolesnikov, D.~Weissenborn, X.~Zhai, T.~Unterthiner, M.~Dehghani, M.~Minderer, G.~Heigold, S.~Gelly, J.~Uszkoreit, and N.~Houlsby, ``An image is worth 16x16 words: Transformers for image recognition at scale,'' in \emph{in Proceedings of International Conference on Learning Representations}, 2021.

\bibitem{9024132}
Y.~Zhang, Y.~Guan, K.~Bian, Y.~Liu, H.~Tuo, L.~Song, and X.~Li, ``{EPASS}360: {QoE}-aware 360-degree video streaming over mobile devices,'' \emph{IEEE Transactions on Mobile Computing}, vol.~20, no.~7, pp. 2338--2353, 2021.

\bibitem{10.1145/3204949.3208139}
E.~J. David, J.~Guti\'{e}rrez, A.~Coutrot, M.~P. Da~Silva, and P.~L. Callet, ``A dataset of head and eye movements for 360° videos,'' in \emph{Proceedings of the 9th ACM Multimedia Systems Conference}, 2018, p. 432–437.

\bibitem{9395242}
M.~F.~R. Rondón, L.~Sassatelli, R.~Aparicio-Pardo, and F.~Precioso, ``Track: A new method from a re-examination of deep architectures for head motion prediction in 360$^{\circ }$ videos,'' \emph{IEEE Transactions on Pattern Analysis and Machine Intelligence}, vol.~44, no.~9, pp. 5681--5699, 2022.

\bibitem{9733647}
F.-Y. Chao, C.~Ozcinar, and A.~Smolic, ``Transformer-based long-term viewport prediction in 360° video: Scanpath is all you need,'' in \emph{Proceedings of IEEE 23rd International Workshop on Multimedia Signal Processing}, 2021, pp. 1--6.

\bibitem{10.1145/3204949.3208123}
D.~Raca, J.~J. Quinlan, A.~H. Zahran, and C.~J. Sreenan, ``Beyond throughput: a {4G LTE} dataset with channel and context metrics,'' in \emph{Proceedings of the ACM Multimedia Systems Conference}, 2018, pp. 460--465.

\bibitem{10.1145/3123266.3123414}
A.~T. Nasrabadi, A.~Mahzari, J.~D. Beshay, and R.~Prakash, ``Adaptive 360-degree video streaming using scalable video coding,'' in \emph{Proceedings of the ACM International Conference on Multimedia}, 2017, pp. 1689--1697.

\bibitem{10.1145/3394171.3413751}
Y.~Mao, L.~Sun, Y.~Liu, and Y.~Wang, ``Low-latency {FoV}-adaptive coding and streaming for interactive 360° video streaming,'' in \emph{Proceedings of the ACM International Conference on Multimedia}, 2020, pp. 3696--3704.

\end{thebibliography}

\begin{IEEEbiography}{Haopeng Wang}{\,} is a Ph.D. candidate with the School of Electrical Engineering and Computer Science, University of Ottawa, Ottawa, ON, Canada. His research interests include multimedia, extended reality, and artificial intelligence. Wang received his M.Sc. degree from the Beijing Institute of Technology, Beijing, China. Contact him at hwang266@uottawa.ca. 
\end{IEEEbiography}

\begin{IEEEbiography}{Haiwei Dong} is a Principal Researcher at Huawei Canada and an Adjunct Professor at the University of Ottawa. His research interests include artificial intelligence, multimedia, metaverse, and robotics. Dong received his Ph.D. degree from Kobe University, Kobe, Japan. He is a Senior Member of IEEE. Contact him at haiwei.dong@ieee.org. 
\end{IEEEbiography}

\begin{IEEEbiography}{Abdulmotaleb El Saddik}
is a distinguished professor with the University of Ottawa, Ottawa, ON, Canada and MBZUAI, UAE. His research interests include the establishment of digital twins to facilitate the well-being of citizens using AI, IoT, AR/VR to allow people to interact in real time with one another as well as with their smart digital representations. He is a fellow of Royal Society of Canada, a fellow of IEEE, an ACM distinguished scientist, and a fellow of the Engineering Institute of Canada and the Canadian Academy of Engineers. Contact him at elsaddik@uottawa.ca.
\end{IEEEbiography}

\end{document}